# Comparison of two photonic sampling mixer architectures based on SOA-MZI for all-optical frequency up-conversion

Dimitrios Kastritsis[1,2,*], Thierry Rampone[1], Kyriakos Zoiros[2], Ammar Sharaiha[1]
[1] Lab-STICC UMR CNRS 6285, École Nationale d'Ingénieurs de Brest (ENIB), Brest, France
[2] Dept. of Electrical and Computer Engineering, Lightwave Communications Research Group, Democritus University of Thrace, Xanthi, Greece

* kastritsis@enib.fr

*Abstract*—An experimental comparison of the conversion gain and harmonic distortion performance between Switching and Modulation architectures of an all-optical photonic sampler mixer up-converter using a Semiconductor Optical Amplifier-based Mach-Zehnder Interferometer (SOA-MZI) is presented. The process of frequency up-conversion from 1 GHz to 9 GHz is evaluated. Because of their different principle of operation, the Switching architecture demonstrates higher positive conversion gain by approximately 6 dB and 8 dB for standard and differential configuration, respectively, while the Modulation architecture achieves lower harmonic distortion up to 8 dB, depending on the modulation index of the 1 GHz signal.

## I. INTRODUCTION

The interdisciplinary field of microwave photonics permits certain functionalities to be performed either in the electrical or in the optical domain [1]. The choice between the two domains depends on the assessment of the advantages and the constraints of the specific electronic or optical module available for this purpose. In this framework, an all-optical frequency mixer is an indispensable element of many applications, mainly because of the perspective to be incorporated into a Radio Frequency (RF) over Fiber (RoF) system and the concomitant benefits that the latter provides, i.e. a large bandwidth, low losses and a high immunity to electromagnetic interference. In particular the use of a sampling mixer allows for simultaneous frequency conversion to multiple frequencies, thus offering an increased system flexibility.

Photonic microwave mixing employing a Semiconductor Optical Amplifier-based Mach-Zehnder Interferometer (SOA-MZI) has been proposed and experimentally demonstrated in [2], where frequency up-conversion from 2.5 GHz to 12.5 GHz and from 2.5 GHz to 22.5 GHz was demonstrated with a maximum conversion gain of 15.33 dB and 5.3 dB, respectively. The sampling process using a SOA-MZI for signal up- and down-conversion has been demonstrated in [3] and [4]. The highest conversion gain was reported in [4], where up-conversion of a 0.5 GHz signal to various frequencies between 8.3 GHz and 39.5 GHz was achieved. The conversion gain in this range was reduced from 15.5 dB to -13.4 dB for standard configuration.

Sampling in SOA-MZI can be obtained through the cross phase modulation (XPM) nonlinear effect induced in SOAs. It can be performed either by using optical train pulses to switch on and off the input optical signal, denoted as 'Switching architecture', or by using input data to modulate the amplitude of the optical pulses, referred to as 'Modulation architecture'. Because XPM has an operation bandwidth limited to a few GHz [5] due to the SOA finite carrier lifetime, the quality of sampling is degraded when using the Switching architecture as the repetition rate of the pulse train gets higher. However, the Modulation architecture based on SOA-MZI has the advantage of supporting optical pulse trains of potentially very high frequencies. By using a SOA-MZI, an all-optical implementation can be achieved as opposed to electro-optical devices utilized before for the same purpose [6].

In this paper, we experimentally investigate and compare the two architectures to implement a photonic sampling mixer using the SOA-MZI as a building module. The Switching architecture has been investigated before for the purposes of signal up- and down- frequency conversions [3, 4]. Both standard and differential configurations are evaluated.

## II. PRINCIPLE OF ALL-OPTICAL SAMPLING MIXER BASED ON SOA-MZI

Fig. 1 shows the conceptual diagram of the switching architecture. An optical pulse train is injected in SOA-MZI port A, while a sinusoidal signal is injected in SOA-MZI port C. This signal will be transmitted at the output depending on the phase difference induced between the MZI upper and lower arms. By injecting a high enough optical power into port A or D, a gain saturation occurs in the upper or lower SOA, thus creating a nonlinear phase shift between the MZI branches. This differential phase shift can be controlled and transformed into an amplitude variation at outputs I and J of SOA-MZI, which acts as an all-optical switch that turns 'off' and 'on' the signal coming from port C. Provided that the switch closes and opens fast enough and the nonlinear phase shift incurred by the optical pulses peak reaches π, the sampled signal is obtained at output J or I, depending on the operating point.

In the spectral domain, the optical pulse train, which acts as the sampling signal at frequency $F_{sa}$, consists of harmonics, at the frequencies $nF_{sa}$, where n is a non-zero integer. The process of sampling a sinusoidal signal can be viewed as a replication of its spectral content on either side of the frequency comb lines spaced equidistantly apart by the frequency of the sinusoidal signal, $F_{IF}$. The desirable output frequency, $nF_{sa} \pm F_{IF}$, can be selected after photo-detection with an electronic bandpass filter.

This standard configuration of the switching architecture can be modified into a differential configuration in which the sampling pulse train is both injected into port A and into port D after a delay.

Fig. 2 shows the conceptual diagram of the modulation architecture. The difference in this architectural approach is that the two input signals, the optical sampling pulse train and the sinusoidal signal, are exchanged between ports A and C compared to the previous architecture. In this way the optical pulses are modulated by the sinusoidal signal to produce the sampled signal at the SOA-MZI output. In the spectral domain, we obtain an identical frequency spectrum as the one of the Switching architecture.

## III. EXPERIMENTAL SETUPS

The operating point of the SOA-MZI (CIP model 40G-2R2-ORP) was selected taking into account many constraints. The SOAs are symmetrically biased with 380 mA current. Choosing a high value for the current improves the dynamic behavior of the SOAs. An Optical Pulse Clock (OPC, Pritel model UOC-E-05-20) source driven by an RF generator RF2 at 10 GHz provides an optical pulse train of 1.3 ps full-width at half-maximum pulses at 1550 nm.

A laser source producing a continuous wave (CW) signal at 1557.4 nm is intensity modulated by an optical Mach-Zehnder Modulator (MZM) driven by an RF generator RF1 at 1 GHz. The electrical power of the photo-detected optical signal injected at the SOA-MZI input port C is denoted as $P_{e,in,dat}$.

*A. Switching Architecture*

The setup of the Switching architecture is shown in Fig. 3. An optical pulse train is injected at port A as a sampling signal and a sinusoidal signal to be sampled is injected at port C.

The sampled signal can be produced at output port J or I of the SOA-MZI, depending on the operating point of the latter. By using Switch 2X1 we can select which output port to observe in a RF Spectrum Analyzer (RFSA) after optical filtering and photo-detection. An optical bandpass filter (OBPF) centered at 1557.4 and of 3 dB bandwidth 0.56 nm rejects any other signal but the sampled one. The combined optical loss of the OBPF and Switch 2X1 is 5.6 dB. The electrical power at the output of SOA-MZI, $P_{e,out,dat}$, is calculated from the RFSA taking into account the aforementioned loss. All the results presented in the following are for Switch 2X1 turned on SOA-MZI port I.

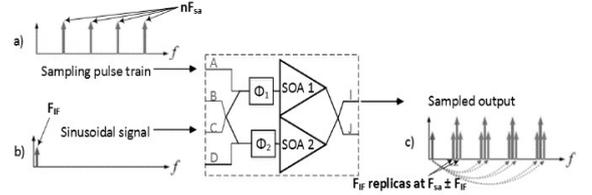

Fig. 1. Conceptual diagram of SOA-MZI sampler with Switching architecture. The vertical arrowed lines indicate the spectral components of a) the sampling pulse train, b) the sinusoidal signal to be sampled and c) the sampled output.

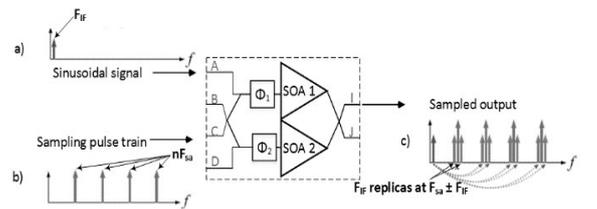

Fig. 2. Conceptual diagram of SOA-MZI sampler with Modulation architecture. The vertical arrowed lines indicate the spectral components of a) the sinusoidal signal to be sampled, b) the sampling pulse train and c) the sampled output.

The differential configuration is implemented by injecting a signal from the OPC directly at port A and a delayed version of it at port D thanks to a variable Optical Delay Line (ODL) so as to produce this delayed version (represented by the dot-dashed line in Fig. 3). The ODL delay is adjusted so that the frequency components of 10, 20, 30 … GHz in the RFSA are as equal as possible when a CW signal is applied at port C, indicating a flat frequency response.

An electrical oscilloscope measures the peak, $V_p$, and mean amplitude values, $V_{mean}$, of the sinusoidal signal at input C in order to calculate its modulation index, $MI = V_p / V_{mean}$.

*B. Modulation Architecture*

Fig.4 shows the scheme implemented for the Modulation architecture. The optical pulse train is inserted in port C, and an intensity-modulated sinusoidal signal is injected at SOA-MZI port A.

The mean output power of the OPC is adjusted by an optical attenuator (Att2) at -15 dBm so as to prevent the generated pulses from saturating the SOA and degrading the performance of the SOA-MZI as a sampler-mixer. The OBPF center frequency is tuned at 1550 nm and selects the sampled signal.

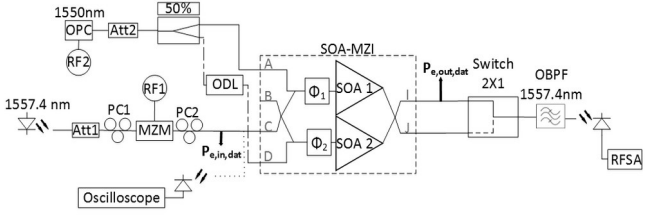

Fig. 3. Standard and Differential configuration setup for Switching architecture. OPC: Optical Pulse Clock generator, Att: Optical Attenuator, PC: Polarization Controller, RF: Radio Frequency generator, MZM: Mach-Zehnder Modulator, ODL: Optical Delay Line, Φ: Phase Shifter, OBPF: Optical Bandpass Filter, RFSA: RF Spectrum Analyzer.

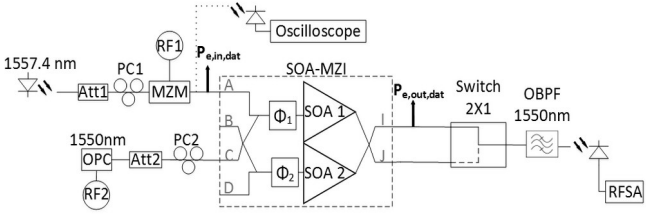

Fig. 4. Configuration setup for Modulation architecture of SOA-MZI.

*C. Performance Metrics*

In order to evaluate the linearity of the sampling mixer, the Harmonic Distortion (H.D.), i.e. HD2 and 3, and the Total Harmonic Distortion (T.H.D.) are used as performance metrics.

At the output of the all-optical mixer in the frequency domain, the 10 GHz frequency component is surrounded by the up-converted signal at 9 GHz and 11 GHz, while the harmonic distortion products appear at 7 GHz and 8 GHz, on the lower sideband, and at 12 GHz and 13 GHz, on the higher sideband. HD2 and HD3 are defined as shown in Fig. 5. T.H.D. is the sum of 7 GHz and 8 GHz distortion frequency components divided by the 9 GHz target frequency component.

The conversion gain from 1 GHz to 9 GHz is measured using the formula:

$$G_{c,up}(dB) = P_{e,out,dat,9GHz}(dBm) - P_{e,in,dat,1GHz}(dBm)$$

Harmonic distortion metrics are very important for RoF systems. Since the SOA-MZI output would be filtered in the electrical domain to acquire the up-converted signal at 9 GHz, lower HD could make possible the use of lower cost and selectivity electronic filter.

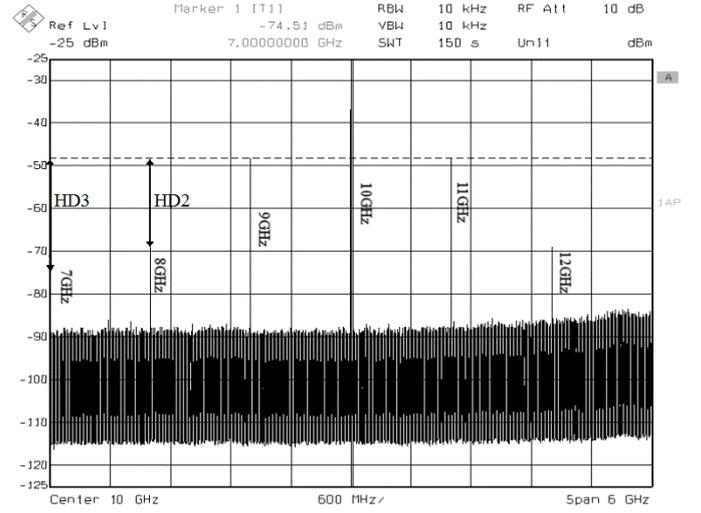

Fig. 5. Electrical spectrum of the sampled signal on the RFSA. Measurement of HD2 and HD3 using lower sideband.

## IV. EXPERIMENTAL RESULTS

The operating point for both architectures was chosen with criterion the minimization of the distortion for a range of modulation indices from 0.2 up to 1.

In order to choose the mean power of the sinusoidal signal for the Switching architecture, a static characterization (identical to the pump-probe static configuration in [3]) was performed for the SOA-MZI, as shown in Fig. 6. In this figure, the derivative of the transmitted power at SOA-MZI output port J is shown as a function of the optical power at input port A. A moving average smoothing of 7-th order was applied on the measured data. A "quasi-linear region" is indicated by the dashed line. In this region the derivative of the transmitted power changes not more than 20% and is assumed to be sufficiently linear to be used during the switching process. The center of this region is almost -10.5 dBm, and the mean optical power at input A is adjusted at this value by the optical attenuator Att2.

For the differential Switching configuration, the operating point corresponds to almost the same optical mean power, -10.7 dBm, which is injected into the input port D.

In order to choose the mean power of the sinusoidal signal for the Modulation architecture, we keep the modulated power $P_{e,in,dat,1GHz}$ constant, sweep the mean power at port A from -20 dBm to -7 dBm, and measure HD2, which is the most significant product, as shown in Fig. 7. The mean power of the lowest HD2 is -14 dBm, and the mean optical power at input A is adjusted at this value by the optical attenuator Att1.

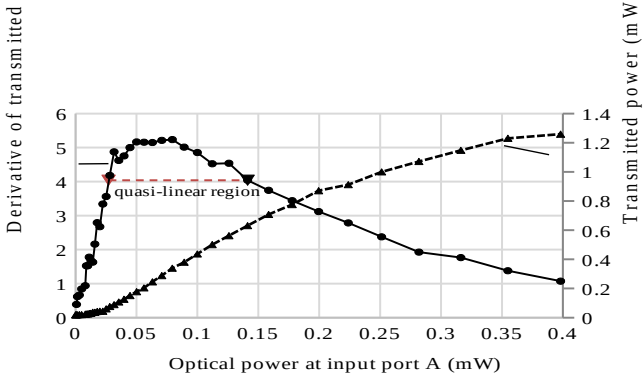

Fig. 6. SOA-MZI static response at port J (right) and its derivative (left) as a function of optical power at input port A.

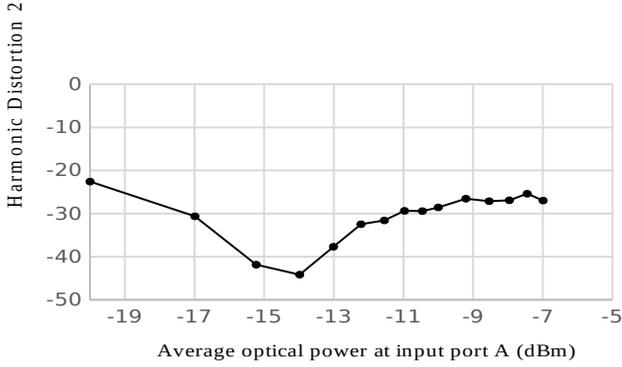

Fig. 7. Harmonic Distortion 2 (HD2) of SOA-MZI as a function of sinusoidal signal mean power for constant modulated power, $P_{e,in,dat,1GHz}$.

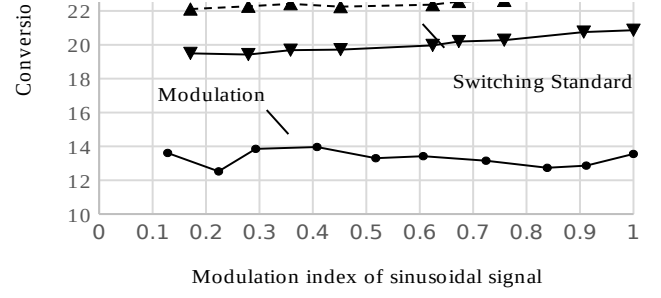

Fig. 8. Comparison of conversion gains between Switching (standard and differential) and Modulation architectures as a function of modulation index of the 1 GHz sinusoidal signal $P_{e,in,dat,1GHz}$.

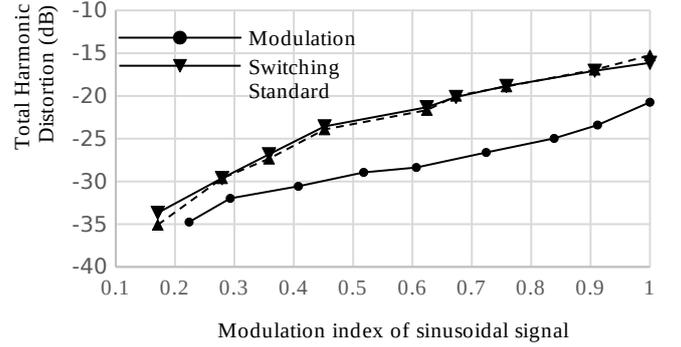

Fig. 9. Comparison of SOA-MZI Total Harmonic Distortion (THD) between Switching (standard and differential) and Modulation architectures as a function of the modulation index of the sinusoidal signal $P_{e,in,dat,1GHz}$.

Fig. 8 8 compares the positive conversion gain obtained with the two architectures as a function of the modulation index of the sinusoidal signal. The highest conversion gain, of about 22.9 dB, is obtained with the Switching architecture using the differential configuration, then for the Switching architecture using the standard configuration of about 20.9 dB and lastly for the Modulation architecture of about 14 dB. For the Switching architecture the conversion gain increases moderately with the modulation index.

Fig. 9 shows a comparison of the THD for the two SOA-MZI architectures. The THD varies significantly for both architectures as the modulation index increases, i.e. by 20 dB for the standard Switching architecture and by 15 dB for the Modulation architecture. The main conclusion from this figure is that the Modulation architecture has a lower THD compared to the Switching architecture. The difference between the two architectures is around 5 dB or more for a modulation index higher than 0.4, while it is as low as 1 dB for a modulation index of 0.2. From this figure we notice that the responses of the standard and differential configurations of Switching architecture are almost identical.

## V. CONCLUSION

In conclusion, we have experimentally investigated and compared two different SOA-MZI-based sampling architectures for signal up-conversion. From this comparison we have found that the Switching architecture can be used in applications that require high conversion gain, while the Modulation architecture can be exploited in applications where the harmonic distortion should be minimum.

ACKNOWLEDGMENT

This research is supported by Brest Métropole (France) through project ÉChantillonnage Optique ultra Rapide à base d'un SOA-MZI pour applications Analogiques Large bande (CHORAAL) and is part of a joint PhD thesis supervision between École Nationale d'Ingénieurs de Brest (ENIB) in France and Democritus University of Thrace (DUTH) in Greece.